\documentclass{jpsj3}

\usepackage{txfonts}
\usepackage{graphicx}
\usepackage{dcolumn}
\usepackage{bm}

\def\lesssim{\ \raise.3ex\hbox{$<$}\kern-0.8em\lower.7ex\hbox{$\sim$}\ }
\def\gesim{\ \raise.3ex\hbox{$>$}\kern-0.8em\lower.7ex\hbox{$\sim$}\ }

\title{Pseudogap regime of a two-dimensional uniform Fermi gas}
\author{Morio Matsumoto$^1$\thanks{moriom@rk.phys.keio.ac.jp}, Ryo Hanai$^2$, Daisuke Inotani$^1$, and Yoji Ohashi$^1$}
\inst{$^1$Department of Physics, Keio University, 3-14-1 Hiyoshi, Kohuku-ku, Yokohama 223-8522, Japan \\
$^2$ Department of Physics, Osaka University, 1-1 Machikaneyama, Toyonaka,  560-0043, Japan.
}

\abst{
We investigate pseudogap phenomena in a two-dimensional Fermi gas. Including pairing fluctuations within a self-consistent $T$-matrix approximation, we determine the pseudogap temperature $T^*$ below which a dip appears in the density of states $\rho(\omega)$ around the Fermi level. Evaluating $T^*$, we identify the pseudogap region in the phase diagram of this system. We find that, while the observed BKT (Berezinskii-Kosterlitz-Thouless) transition temperature $T^{\rm exp}_{\rm BKT}$ in a $^6$Li Fermi gas is in the pseudogap regime, the detailed pseudogap structure in $\rho(\omega)$ at $T^{\rm exp}_{\rm BKT}$ still differs from a fully-gapped one, indicating the importance of amplitude fluctuations in the Cooper channel there. Since the observed $T^{\rm exp}_{\rm BKT}$ in the weak-coupling regime cannot be explained by the recent BKT theory which only includes phase fluctuations, our results may provide a hint about how to improve this BKT theory. Although $\rho(\omega)$ has not been measured in this system, we show that the assessment of our results is still possible by using the {\it observable} Tan's contact.
}

\begin{document}
\maketitle
\section{Introduction}
\par
In cold Fermi gas physics, there are two ways to tune pairing fluctuations. One is to use a Feshbach resonance\cite{Timmermans,Chin}, where pairing fluctuations are enhanced by increasing the strength of a pairing interaction associated with a Feshbach resonance. The other is lowering the system dimension by introducing an optical lattice\cite{Morsch,Bloch} to a three-dimensional Fermi gas. The latter approach uses the fact that pairing fluctuations are enhanced by the low-dimensionality of the system. In condensed matter physics, it is not so easy to tune the interaction strength, as well as the system dimension. Thus, having these techniques is an advantage of cold Fermi gas physics, especially in systematically studying strong-coupling physics.
\par
In this paper, we theoretically investigate the pseudogap phenomenon in an ultracold Fermi gas, when pairing fluctuations are enhanced by lowering the system dimension down to two. Pseudogap is a typical many-body phenomenon associated with strong pairing fluctuations\cite{Perali,Levin,Tsuchiya,Watanabe1}, and is characterized by the appearance of a dip in the single-particle density of states $\rho(\omega)$ (DOS) around the Fermi level in the normal state. When one simply describes pairing fluctuations as the repeat of the formation and dissociation of preformed Cooper pairs, the pseudogap is physically interpreted as a result of the ``binding energy" of preformed Cooper pairs. In this paper, we calculate the pseudogapped DOS in a two-dimensional uniform Fermi gas using a strong-coupling theory (which we specify soon later). From the temperature dependence of the pseudogap, we determine the pseudogap temperature $T^*$ as the temperature below which a dip appears in $\rho(\omega)$. Using $T^*$, we identify the pseudogap regime (where the pseudogap appears in $\rho(\omega)$) in the phase diagram of a two-dimensional Fermi gas with respect to the temperature and the interaction strength. We briefly note that a two-dimensional Fermi gas has recently attracted much attention both theoretically\cite{Botelho,Tempere,Pietilla,Watanabe,Bauer,Mulkerin,Marsiglio,Matsumoto1,Salasnich,Mulkerin1} and experimentally\cite{Martiyanov,Feld,Frohlich,Sommer,Makhalov,Fenech,Ries,Murthy2}, especially since the observation of the Berezinskii-Kosterlitz-Thouless(BKT) transition\cite{Berezinskii,Berezinskii1,Kosterlitz,Kosterlitz1} in a $^6$Li Fermi gas\cite{Ries,Murthy2}. 
\par
The importance of studying two-dimensional pseudogap phenomena is related to the current debate for this phenomenon in a three-dimensional Fermi gas. In the three-dimensional case, the ``Feshbach-resonance approach" has extensively been used to study the BCS (Bardeen-Cooper-Schrieffer)-BEC (Bose-Einstein condensation) crossover phenomenon\cite{Chin,Regal,Zwierlein,Kinast,Bartenstein,Bloch,Ketterle,Zwerger}, where the character of a Fermi superfluid gradually deviates from the weak-coupling BCS-type with increasing the strength of a pairing interaction\cite{Eagles,Leggett,NSR,Melo,Randeria,Ohashi2002,Levin2005,Gurarie,Giorgini}. In the intermediate coupling regime (BCS-BEC crossover region), system properties are dominated by strong pairing fluctuations, where the pseudogap has been expected\cite{Tsuchiya,Watanabe1,Chen,Hu,Mueller,Bulgac,Jarrell,Bulgac2,Ota}. However, while the recent photoemission-type experiments on $^{40}$K Fermi gases\cite{Jin1,Jin2,Jin3,Jin4} agree with the pseudogap scenario\cite{Tsuchiya,Watanabe,Chen,Hu,Mueller,Bulgac,Jarrell,Bulgac2,Ota}, it has been shown that the observed pressure\cite{Salomon1}, as well as spin polarization rate\cite{Salomon2}, can be explained by the normal Fermi liquid theory {\it without} assuming the pseudogap. Thus, further studies are necessary to clarify whether or not pairing fluctuations really cause the pseudogap phenomenon. The mechanism of the pseudogap originating from strong pairing fluctuations is sometimes referred to as the preformed pair scenario\cite{Perali,Randeria1992,Singer,Rohe,Yanase}, and has been discussed as a candidate for the origin of the pseudogap in the under-doped regime of high-$T_{\rm c}$ cuprates\cite{Kadowaki,Shen,Fischer}. Thus, once the validity of this scenario is confirmed in cold Fermi gas physics, it would also make an impact on condensed matter physics.
\par
At a glance, using a Feshbach resonance seems more effective than the optical-lattice method for the study of pseudogap phenomenon, because the former can directly adjust the interaction strength. However, we recall that pairing fluctuations in one- and two-dimensional systems are very strong, to {\it completely} destroy the superfluid long-range order (Hohenberg-Mermin-Wagner's theorem)\cite{Hohenberg,Mermin}. In the three-dimensional case, on the other hand, although the superfluid phase transition temperature $T_{\rm c}$ is suppressed by pairing fluctuations (compared to the mean-field $T_{\rm c}$)\cite{NSR,Melo,Randeria}, the superfluid long-range order itself is always realized in the whole BCS-BEC crossover region. In this sense, lowering the system dimension is more effective than simply increasing the interaction strength, in order to enhance pairing fluctuations. Since the above-mentioned controversial situation implies that three-dimensional pairing fluctuations may not be strong enough to clearly observe the pseudogap phenomenon even in the BCS-BEC crossover region, the further enhancement of  pairing fluctuations by lowering the system dimension is a promising idea to resolve this debate. We briefly note that, even in the two-dimensional case, one can still use a Feshbach resonance\cite{Martiyanov,Feld,Frohlich,Sommer,Makhalov,Fenech,Ries,Murthy2}.
\par
Besides the pseudogap problem, the detailed pseudogap structure in $\rho(\omega)$ may also provide useful information about how to theoretically deal with the BKT transition in a two-dimensional Fermi gas. It is known that some strong coupling theories, such as the Gaussian fluctuation theory\cite{NSR,Melo,Randeria,Ohashi2002}, non-selfconsistent $T$-matrix approximation (TMA)\cite{Perali,Tsuchiya}, as well as their extended versions\cite{Haussmann,Haussmann2}, that have extensively been used to successfully explain the BCS-BEC crossover phenomenon in a three-dimensional Fermi gas, cannot describe the BKT transition, when simply applied to the two-dimensional case. To overcome this problem, a useful approach has been proposed\cite{Botelho,Tempere}, where phase fluctuations of the superfluid order parameter $\Delta$ are only taken into account, under the assumption that amplitude fluctuations of $\Delta$ can be ignored, to give a fixed value of $|\Delta|$. In this theory, with help of the KT-Nelson formula\cite{Nelson}, the BKT transition temperature $T^{\rm th}_{\rm BKT}$ can be evaluated in the wide interaction regime. This approaches uses the fact that the BKT transition belongs to the 2D-XY universality class, that is, the amplitude and phase of $\Delta$ correspond to the length and the direction of spin in the ordinary 2D-XY model, respectively. However, it is still unclear whether or not the assumption of the fixed $|\Delta|$ (or ignoring amplitude fluctuations) is always valid for the whole interaction regime. Indeed, while the calculated $T^{\rm th}_{\rm BKT}$\cite{Botelho,Tempere} agrees with the recent experiment on a two-dimensional $^6$Li Fermi gas\cite{Ries,Murthy2} in the strong-coupling regime, one clearly sees discrepancy between the two in the weak-coupling regime, implying the necessity of improving this theory. Since the fixed $|\Delta|$ gives a BCS-state-like fully gapped DOS, one can check the validity of this assumption from the detailed pseudogap structure near $T^{\rm th}_{\rm BKT}$. 
\par
In this paper, we employ a self-consistent $T$-matrix approximation (SCTMA)\cite{Haussmann,Haussmann2}, to include two-dimensional pairing fluctuations. Although a non-selfconsistent $T$-matrix approximation (TMA) has extensively been used to study the pseudogap phenomenon in the three-dimensional case, we do not take this approach, because TMA is known to unphysically give large pseudogap size at low temperatures, when applied to a two-dimensional Fermi gas, even in the weak-coupling regime\cite{Marsiglio,MatsumotoJLTP}. We briefly note that SCTMA has recently been applied to the weak-coupling regime of a two-dimensional Fermi gas\cite{Bauer,Mulkerin}. 
\par
So far, DOS has not been measured in a two-dimensional Fermi gas, so that we cannot completely assess the validity of our SCTMA approach, by directly comparing the calculated DOS with experimental data. However, as an alternative strategy, we show that the pseudogap size in SCTMA does not contradict with the {\it observed} Tan's contact $C$\cite{Tan} in a two-dimensional $^6$Li Fermi gas\cite{2DTan40K}.
\par
This paper is organized as follows. In Sec. 2, we explain our formulation. In Sec. 3, we show the density of states $\rho(\omega)$, as well as the spectral weight $A({\bm p},\omega)$. From the temperature dependence of the pseudogap in $\rho(\omega)$, we determine the pseudogap temperature $T^*$, to identify the pseudogap regime in the phase diagram with respect to the temperature and the interaction strength. We also examine the detailed pseudogap structure near $T_{\rm BKT}$ which has been evaluated by a previous theory assuming a fixed $|\Delta|$, to discuss the importance of amplitude fluctuations of the superfluid order parameter there. In Sec. 4, we assess our SCTMA results by using the Tan's contact. Throughout this paper, we take $\hbar=k_{\rm B}=1$, and the two-dimensional system area is taken to be unity, for simplicity.
\par
\par
\section{Formulation}
\par
We consider a two-dimensional uniform Fermi gas, described by the Hamiltonian,
\begin{eqnarray}
H
&=&
\sum_{{\bm p},\sigma}\xi_{\bm p}c^\dagger_{{\bm p},\sigma}c_{{\bm p},\sigma}
\nonumber
\\
&-&
U\sum_{{\bm p},{\bm p}',{\bm q}}
c^\dagger_{{\bm p}+{\bm q}/2,\uparrow}
c^\dagger_{-{\bm p}+{\bm q}/2,\downarrow}
c_{-{\bm p}'+{\bm q}/2,\downarrow}
c_{{\bm p}'+{\bm q}/2,\uparrow},
\label{eq1}
\end{eqnarray}
where $c^\dagger_{{\bm p},\sigma}$ is the creation operator of a Fermi atom with the two-dimensional momentum ${\bm p}=(p_x, p_y)$ and pseudospin $\sigma=\uparrow,\downarrow$, describing two atomic hyperfine states. $\xi_{\bm p}=\varepsilon_{\bm p}-\mu={\bm p}^2/(2m)-\mu$ is the kinetic energy, measured from the Fermi chemical potential $\mu$, where $m$ is an atomic mass. $-U~(<0)$ is a tunable pairing interaction associated with a Feshbach resonance. As usual, we measure the interacting strength in terms of the $s$-wave scattering length $a_{2{\rm D}}$, which is related to the bare interaction $-U$ as\cite{Morgan},
\begin{equation}
{2\pi \over m}
\left[\ln(k_{\rm F}a_{2{\rm D}})\right]^{-1}
=
{U \over 1-U\sum_{p\ge k_{\rm F}}{1 \over 2\varepsilon_{\bm p}}}.
\label{eq2}
\end{equation}
Here, $k_{\rm F}=\sqrt{2\pi N}$ is the Fermi momentum, where $N$ is the total number of Fermi atoms. In this scale, the weak- and strong-coupling regime are, respectively, characterized as $\ln{(k_{\rm F}a_{2{\rm D}})} \gesim 1$ and $\ln{(k_{\rm F}a_{2{\rm D}})} \lesssim -1$. The region between the two ($-1\lesssim \ln{(k_{\rm F}a_{2{\rm D}})} \lesssim 1$) is the intermediate coupling regime. 
\par
We briefly note that the present two-dimensional system always has a bound state in the two-particle limit, irrespective of the interaction strength. The two-body binding energy is given by\cite{Petrov,Levinsen,noteEb},
\begin{equation}
E_{\rm bind}={1 \over ma_{\rm 2D}^2}.
\label{eq2b}
\end{equation}
This is different from the three-dimensional case, where a two-body bound molecule is only formed in the strong-coupling regime where the three-dimensional $s$-wave scattering length $a_s$ is positive\cite{Randeria}.
\par
\begin{figure}
\begin{center}
\includegraphics[width=7.5cm]{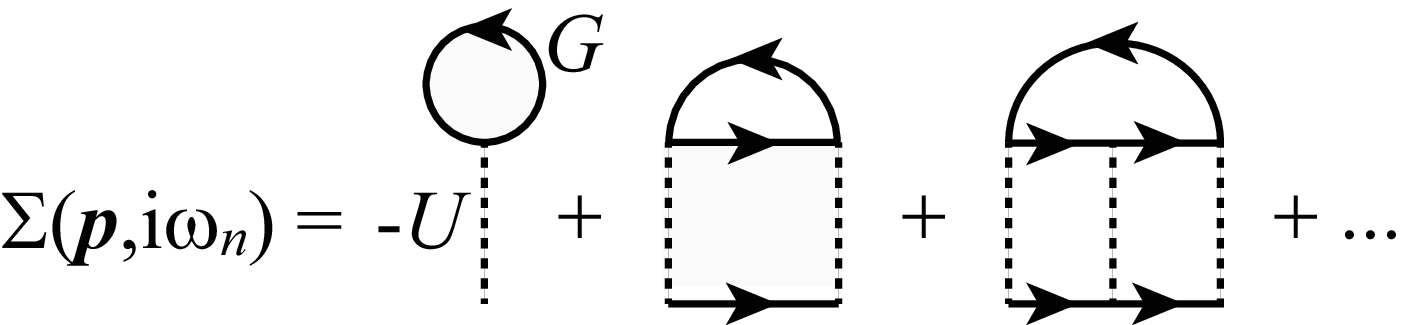}
\caption{Self-energy $\Sigma({\bm p},i\omega_n)$ in the self-consistent $T$-matrix approximation (SCTMA). The solid line and the dashed line denote the dressed single-particle thermal Green's function $G$ in Eq. (\ref{eq3}), and the pairing interaction $-U$, respectively.}
\label{fig1}       
\end{center}
\end{figure}
\par
In the self-consistent $T$-matrix approximation (SCTMA)\cite{Haussmann,Bauer, Mulkerin}, strong-coupling corrections to single-particle excitations are described by the self-energy $\Sigma({\bm p},i\omega_n)$ in the single-particle thermal Green's function,
\begin{equation}
G({\bm p},i\omega_n)=
{1 \over i\omega_n-\xi_{\bm p}-\Sigma({\bm p},i\omega_n)},
\label{eq3}
\end{equation}
where $\omega_n$ is the fermion Matsubara frequency. The SCTMA self-energy $\Sigma(\bm{p},i\omega_{n})$ is diagrammatically described as Fig. \ref{fig1}, which gives
\begin{equation}
\Sigma({\bm p},i\omega_n)
=T\sum_{{\bm q},i\nu_n}
\Gamma({\bm q},i\nu_n)G({\bm q}-{\bm p},i\nu_n-i\omega_n).
\label{eq4}
\end{equation}
Here, $\nu_n$ is the boson Matsubara frequency, and 
\begin{equation}
\Gamma({\bm q},i\nu_n)=-{U \over 1-U\Pi({\bm q},i\nu_n)}
\label{eq5}
\end{equation}
is the particle-particle scattering matrix, where 
\begin{equation}
\Pi({\bm q},i\nu_n)=T\sum_{{\bm p},i\omega_n}
G({\bm p}+{\bm q}/2,i\omega_{n}+i\nu_n)
G(-{\bm p}+{\bm q}/2,-i\omega_{n})
\label{eq6}
\end{equation}
is a pair-correlation function, describing fluctuations in the Cooper channel.
\par
\begin{figure}
\begin{center}
\includegraphics[width=7.5cm]{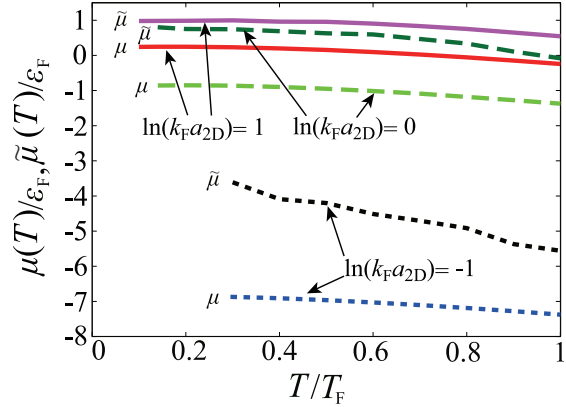}
\caption{(Color online) Calculated Fermi chemical potential $\mu(T)$, as well as the renormalized chemical potential ${\tilde \mu}(T)$ in Eq. (\ref{eq12}), in a two-dimensional Fermi gas. At each interaction strength, ${\tilde \mu}(T)$ is larger than $\mu(T)$. The interaction strength is measured in terms of $\ln(k_{\rm F}a_{\rm 2D})$, where $a_{\rm 2D}$ is the $s$-wave scattering length in Eq. (\ref{eq2}) and $k_{\rm F}$ is the Fermi momentum. $T_{\rm F}$ and $\varepsilon_{\rm F}$ are the Fermi temperature and the Fermi energy, respectively.  Because of computational problems, we could not calculate $\mu$ down to $T=0$. The lowest temperature at each interaction strength is nothing to do with the BKT transition temperature.
}
\label{fig2}       
\end{center}
\end{figure}
\par
The Fermi chemical potential $\mu$ is determined from the equation for the total number $N$ of Fermi atoms,
\begin{equation}
N=2T\sum_{{\bm p},i\omega_n}G({\bm p},i\omega_n). 
\label{eq7}
\end{equation}
We show the calculated $\mu$ in Fig. \ref{fig2}.
\par
Using $\mu$ in Fig. \ref{fig2}, we calculate the single-particle spectral weight,
\begin{equation}
A({\bm p},\omega)=-{1 \over \pi}
{\rm Im}G({\bm p},i\omega_n\to \omega+i\delta),
\label{eq8}
\end{equation}
as well as the single-particle density of states (DOS),
\begin{equation}
\rho(\omega)=\sum_{\bm p}A({\bm p},\omega),
\label{eq.8b}
\end{equation}
where $\delta$ is an infinitesimally small positive number. In this paper, we employ the Pad\'e approximation\cite{Vidberg}, to numerically carry out the analytic continuation ($i\omega_n\to\omega+i\delta$) in Eq. (\ref{eq8}).
\par
For later convenience, we briefly note that the non-self-consistent $T$-matrix approximation (TMA) is given by simply replacing the {\it dressed} Green's function $G({\bm p},i\omega_n)$ in Eqs. (\ref{eq4}) and (\ref{eq6}) with the {\it bare} one,
\begin{equation}
G_0({\bm p},i\omega_n)=
{1 \over i\omega_n-\xi_{\bm p}}.
\label{eq6b}
\end{equation}
\par
As mentioned previously, SCTMA and TMA cannot describe the superfluid (BKT) phase below $T_{\rm BKT}$. These strong-coupling theories determine the superfluid instability from the Thouless criterion\cite{Thouless}, stating the appearance of gapless Goldstone collective mode associated with the broken $U(1)$ gauge symmetry. This condition is achieved, when the particle-particle scattering matrix in Eq. (\ref{eq5}) has a pole at ${\bm q}=\nu_n=0$, that is, 
\begin{equation}
\Gamma^{-1}(0,0)=0. 
\label{eq8c}
\end{equation}
However, in the two-dimensional case, Eq. (\ref{eq8c}) cannot simultaneously be satisfied with the required number equation (\ref{eq7}), because the latter diverges when the former condition is satisfied. This is a direct consequence of the Hohenberg's theorem\cite{Hohenberg}, stating the vanishing superfluid long-range order in one- and two-dimensional systems, due to the enhanced superfluid fluctuations by the low-dimensionality. In the context of two-dimensional BCS-BEC crossover physics, this problem has also been discussed within the framework of the Gaussian fluctuation theory\cite{Varma,Tokumitu}. 
\par
\begin{figure}[t!]
\begin{center}
\includegraphics[width=7cm]{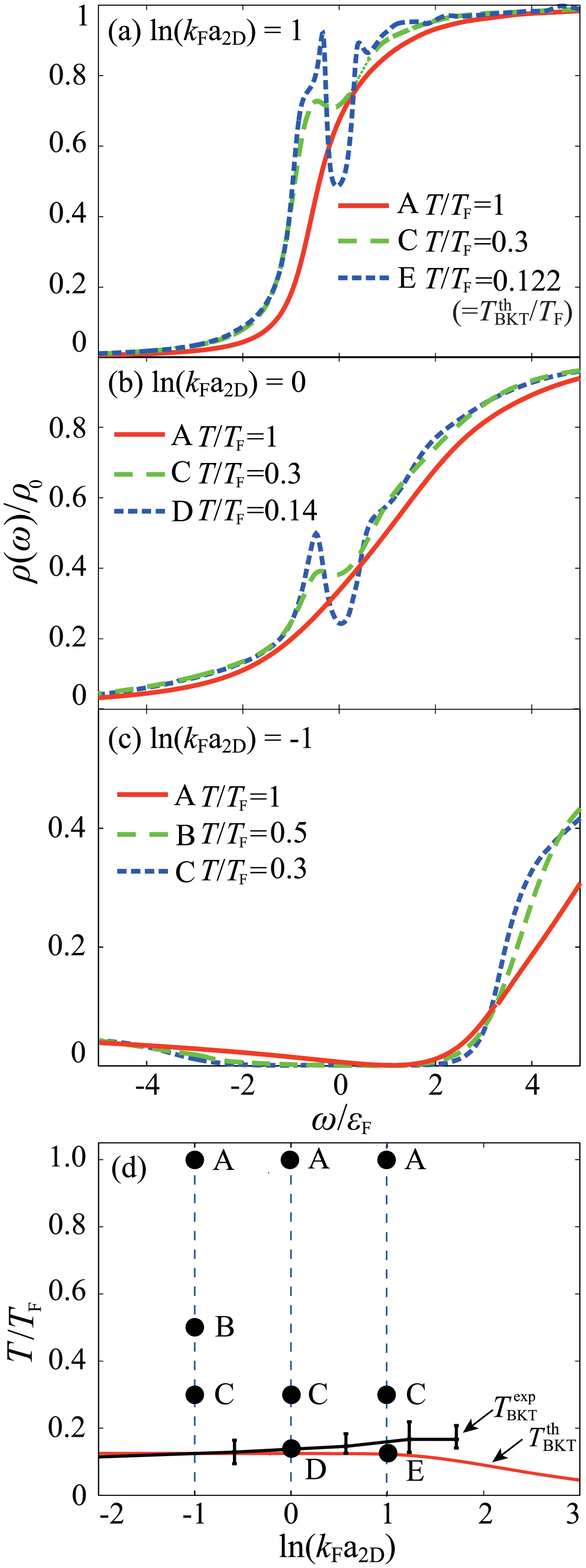}
\caption{(Color online) SCTMA single-particle density of states $\rho(\omega)$ in the normal state of a two-dimensional Fermi gas. $\rho_0=m/(2\pi)$ is DOS in a free Fermi gas. (a) Weak-coupling case ($\ln{(k_{\rm F}a_{2{\rm D}})}=1$). (b) Intermediate coupling case ($\ln{(k_{\rm F}a_{2{\rm D}})}=0$). (c) Strong-coupling case ($\ln{(k_{\rm F}a_{2{\rm D}})}=-1$). Panel (d) shows the position of each result in the phase diagram with respect to the temperature and the interaction strength (where $T_{\rm F}$ is the Fermi temperature). In this figure, $T_{\rm BKT}^{\rm th}$ is the calculated BKT transition temperature by a BKT theory assuming a fixed $|\Delta|$\cite{Botelho,Tempere}. $T_{\rm BKT}^{\rm exp}$ is the observed one in a two-dimensional $^6$Li Fermi gas\cite{Ries}.
}
\label{fig3}       
\end{center}
\end{figure}
\par
\begin{figure}[t!]
\begin{center}
\includegraphics[width=7.5cm]{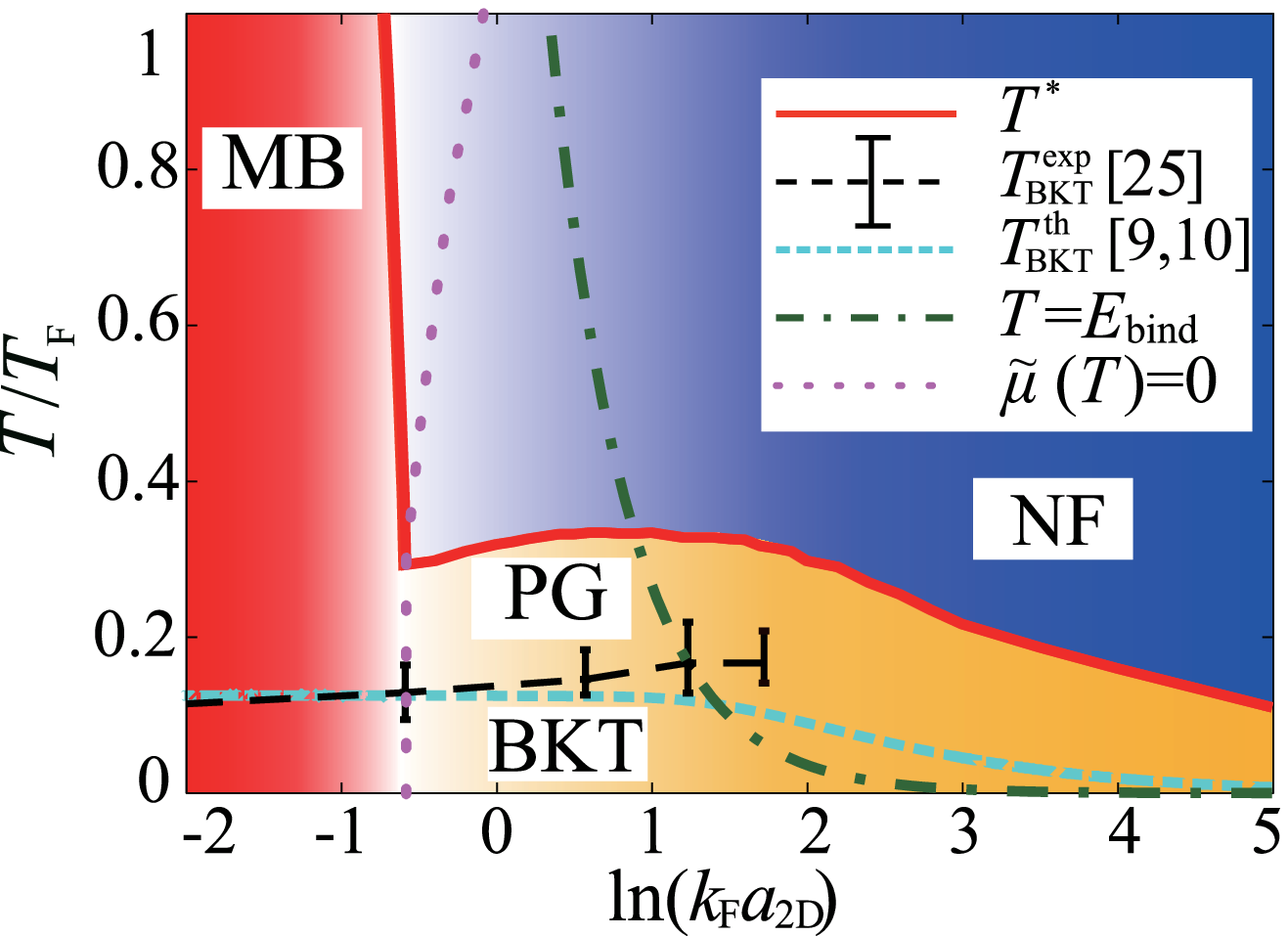}
\caption{ (Color online) Phase diagram of a two-dimensional uniform Fermi gas in terms of the temperature and the interaction strength. The pseudogap temperature $T^*$ is determined as the temperature below which a dip appears in $\rho(\omega\sim 0)$. The two BKT transition temperatures, $T_{\rm BKT}^{\rm exp}$\cite{Ries} and $T_{\rm BKT}^{\rm th}$\cite{Botelho,Tempere}, are the same as those in Fig. \ref{fig3}(d). The region between $T^*$ and $T_{\rm BKT}^{\rm th}$ is the pseudogap regime (PG). The region above $T^*$ is the ``normal Fermi-gas regime (NF)" with no pseudogap. On the left side of the dotted line (MB), the renormalized chemical potential ${\tilde \mu}$ in Eq. (\ref{eq12}) is negative, so that the system properties is close to those of a Bose gas of two-body bound molecules. We briefly note that the phase transition only occurs at $T_{\rm BKT}^{\rm th}$ and $T_{\rm BKT}^{\rm exp}$. The other boundaries, such as $T^*$, are crossover temperatures without being accompanied by any phase transition.}
\label{fig4}       
\end{center}
\end{figure}
\par
\section{Pseudogap regime in a two-dimensional Fermi gas}
\par
Figure \ref{fig3} shows DOS $\rho(\omega)$ in a two-dimensional uniform Fermi gas. In the weak-coupling case (panel (a)), $\rho(\omega)$ at the Fermi temperature $T_{\rm F}$ monotonically increases with increasing $\omega$, to approach the  value $\rho_0=m/(2\pi)$ in the high-energy region. Since DOS in a two-dimensional free Fermi gas has the step-functional energy dependence as
\begin{equation}
\rho_0(\omega)={m \over 2\pi}\Theta(\omega+\mu)
\label{eq8d}
\end{equation}
(where $\Theta(x)$ is the step function), the broadening of this step structure seen in Fig. \ref{fig3}(a) at $T=T_{\rm F}$ is due to particle-particle scatterings (giving finite lifetime of Fermi quasi-particles). Indeed, this broadening at $T=T_{\rm F}$ becomes more remarkable in the intermediate coupling case shown in Fig. \ref{fig3}(b).
\par
In addition to this broadening effect, Figs. \ref{fig3}(a) and (b) also show that the pairing interaction (which induces pairing fluctuations) also causes the expected pseudogap phenomenon at low temperatures ($T\lesssim 0.3T_{\rm F}$). When we introduce the pseudogap temperature $T^*$ as the temperature below which a dip appears in $\rho(\omega)$ around $\omega=0$, we can conveniently identify the region where the pseudogap appears, in the phase diagram of a two-dimensional Fermi gas, as shown in Fig. \ref{fig4}. In this figure, since SCTMA cannot describe the BKT transition, we draw $T^{\rm th}_{\rm BKT}$ which has been evaluated by the recent BKT theory assuming a fixed $|\Delta|$\cite{Botelho,Tempere}. In this phase diagram, the region between $T^*$ and $T^{\rm th}_{\rm BKT}$ is regarded as the pseudogap regime (PG). We emphasize that, while $T^{\rm th}_{\rm BKT}$ is a phase transition temperature, the pseudogap temperature $T^*$ is a crossover temperature, without being accompanied by any phase transition. 
\par
In the strong-coupling regime shown in Fig. \ref{fig3}(c), a dip structure is already seen at $T=T_{\rm F}$. In addition, DOS exhibits an almost fully gapped structure, when $T\lesssim 0.5T_{\rm F}$. In this regime, the {\it renormalized} chemical potential ${\tilde \mu}$ involving many-body corrections, which is determined from the equation,
\begin{equation}
-{\tilde \mu}= -\mu+{\rm{Re}}[\Sigma({\bm p}=0, \omega_{+}
=-{\tilde \mu}+i\delta)],
\label{eq12}
\end{equation}
is negative (see Fig. \ref{fig2}). Thus, as well known in BCS-BEC crossover physics\cite{Eagles,Leggett,NSR,Melo,Randeria,Ohashi2002}, system properties in this regime would be close to those of a Bose gas of two-body bound molecules, rather than an atomic Fermi gas\cite{note}. To emphasize this, we identify the region with ${\tilde \mu}\le 0$ as the ``molecular Bose gas regime" (MB) in Fig. \ref{fig4}. In this strong-coupling regime, the pseudogap temperature $T^*$, as well as the pseudogap width, are considered to be directly related to the binding energy $E_{\rm bind}$ of a two-body bound molecule in Eq (\ref{eq2b}). Indeed, Fig. \ref{fig4} shows that $T^*$ rapidly increases with increasing the interaction strength in this regime, reflecting the rapid increase of $E_{\rm bind}$. 
\par
\begin{figure}[t!]
\begin{center}
\includegraphics[width=7.5cm]{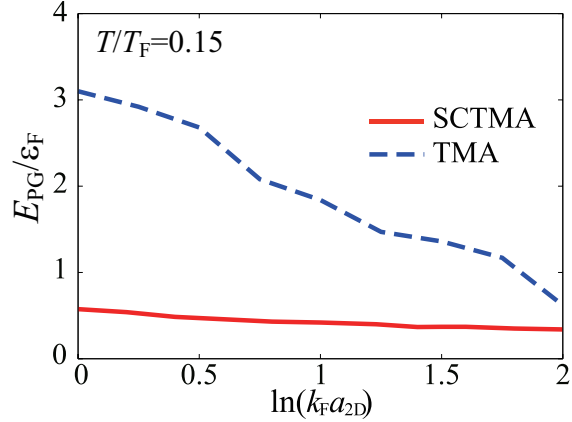}
\caption{(Color online) Evaluated pseudogap size $E_{\rm PG}$ in the weak-coupling side ($\ln(k_{\rm F}a_{\rm 2D})\ge 0$) of a two-dimensional Fermi gas at $T/T_{\rm F}=0.15$. $E_{\rm PG}$ is defined as the energy difference between the dip position and the lower peak position in $\rho(\omega)$ in the pseudogap regime.}
\label{fig5}       
\end{center}
\end{figure}
\par
Figure \ref{fig4} also shows that the pseudogap regime widely exists in the weak-coupling regime ($\ln(k_{\rm F}a_{\rm 2D})\gesim 1)$, compared to the three-dimensional case, where the pseudogap regime is almost absent in the BCS regime (($k_{\rm F}a_s)^{-1}\lesssim -1$)\cite{Tsuchiya}. This is because the two-dimensionality of the system enhances pairing fluctuations, as well as the pseudogap phenomenon, as expected. 
\par
However, we also see in Fig. \ref{fig4} that $T^*$ is rather insensitive to the interaction strength, when $0\lesssim \ln(k_{\rm F}a_{\rm 2D})\lesssim 2$, in spite of the fact that pairing fluctuations become strong with increasing the interaction strength there. In addition, defining the pseudogap size $E_{\rm PG}$ as the energy difference between the dip position and the lower peak position in the pseudogapped DOS\cite{noteZ}, we find in Fig. \ref{fig5} that $E_{\rm PG}$ at $T=0.15T_{\rm F}$ only weakly depends on the interaction strength. These indicate that the increase of the interaction strength does not necessarily promote the pseudogap phenomenon. 
\par
\begin{figure}[t!]
\begin{center}
\includegraphics[width=7.5cm]{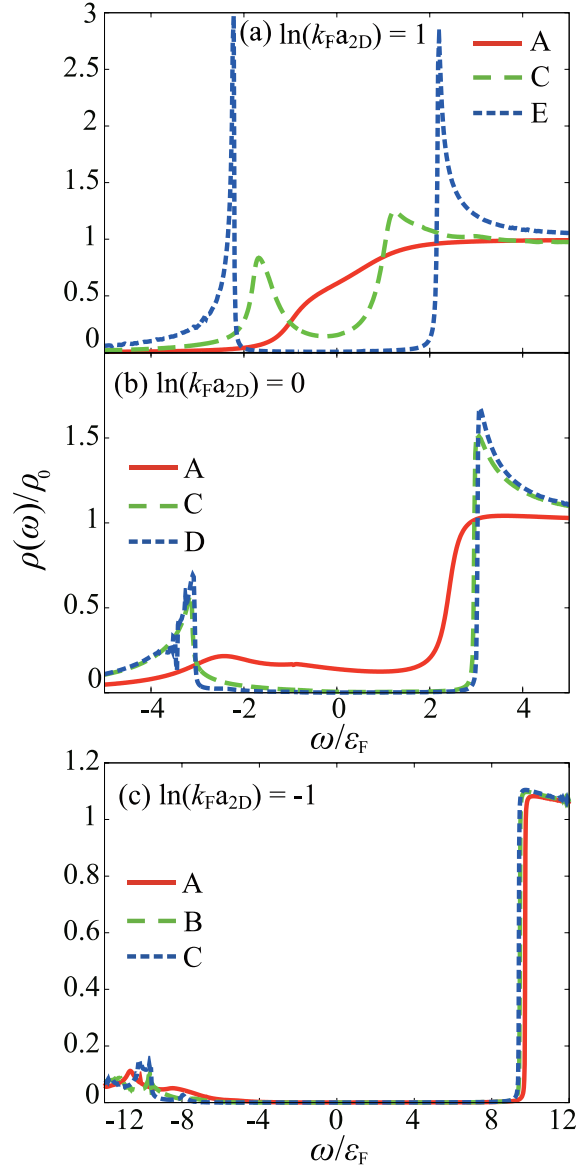}
\caption{The same plots as in Fig. \ref{fig3}, in the non-self-consistent $T$-matrix approximation (TMA).}
\label{fig6}       
\end{center}
\end{figure}
\par
\begin{figure}[t!]
\begin{center}
\includegraphics[width=8cm]{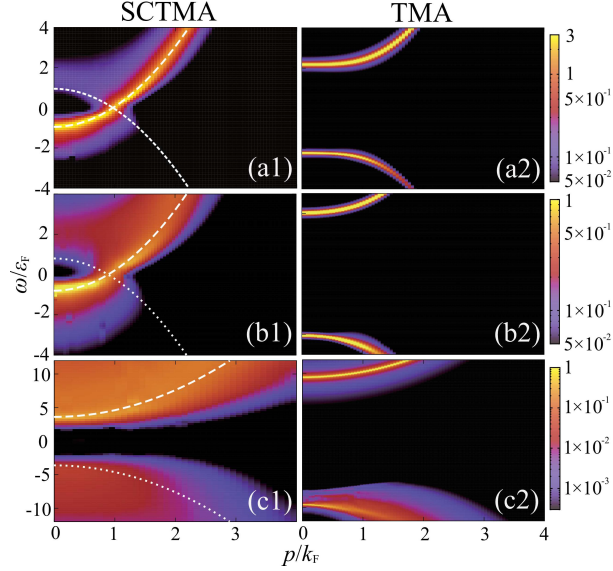}
\caption{(Color online) Calculated intensity of single-particle spectral weight $A({\bm p},\omega)$ in SCTMA (left panels), as well as in TMA (right panels). (a1) and (a2): $\ln(k_{\rm F}a_{\rm 2D})=1$ at $T/T_{\rm F}=0.122$ (``E" in Fig. \ref{fig3}(d)). (b1) and (b2): $\ln(k_{\rm F}a_{\rm 2D})=0$ at $T/T_{\rm F}=0.14$ (``D" in Fig. \ref{fig3}(d)). (c1) and (c2): $\ln(k_{\rm F}a_{\rm 2D})=-1$ at $T/T_{\rm F}=0.3$ (``C" in Fig. \ref{fig3}(d)). In the left figures, the dashed and dotted lines show the particle dispersion ($\omega=\varepsilon_{\bm p}-{\tilde \mu}$) and the hole dispersion ($\omega=-[\varepsilon_{\bm p}-{\tilde \mu}]$), respectively. The intensity is normalized by $\varepsilon_{\rm F}^{-1}$. 
}
\label{fig7}       
\end{center}
\end{figure}

To explain this in more detail, we point out that the non-selfconsistent $T$-matrix approximation (TMA) gives a clearer pseudogap than SCTMA, as shown in Fig. \ref{fig6}. In addition, the growth of the TMA pseudogap size $E_{\rm PG}$ with increasing the interaction strength is also more remarkable than the SCTMA case (see Fig. \ref{fig5}). Thus, in TMA, the pairing interaction simply contribute to the pseudogap phenomenon in a positive manner. In this regard, we note that TMA is given by replacing the SCTMA dressed Green's function $G({\bm p},i\omega_n)$ in the self-energy $\Sigma({\bm p},i\omega_n)$ in Eq. (\ref{eq4}) with the bare one $G_0({\bm p},i\omega_n)$ in Eq. (\ref{eq6b}). The resulting TMA particle-particle scattering matrix $\Gamma({\bm q},i\nu_n)$ (which physically describes pairing fluctuations) only consists of {\it stable} Fermi atoms with infinite lifetime. On the other hand, $\Gamma({\bm q}i\nu_n)$ in SCTMA uses the dressed Green's function $G({\bm p},i\omega_n)$, which takes into account finite quasi-particle lifetime by particle-particle scatterings. Indeed, as shown in Figs. \ref{fig7}(a1)-(c1), the SCTMA spectral weight $A({\bm p},\omega)$ exhibits {\it broad} single-particle dispersion along $\omega=\varepsilon_{\bm p}-{\tilde \mu}$ in the pseudogap regime. (Note that the bare Green's function gives a $\delta$-functional peak line along $\omega=\varepsilon_{\bm p}-\mu$ in $A({\bm p},\omega)$.) As a result, SCTMA includes the situation that one of two Fermi atoms that are repeating the formation and dissociation of a preformed Cooper pair is scattered into another state, causing the suppression of pairing fluctuations compared to the TMA case\cite{noteW}. This explains the difference between Figs. \ref{fig3} (SCTMA) and \ref{fig6}(TMA). 
\par
In addition, the quasi-particle lifetime would be shorter for a stronger interaction, which tends to weaken the enhancement of pairing fluctuations by the same pairing interaction. This also explains the weaker interaction dependence of the pseudogap size $E_{\rm PG}$ in SCTMA than that in TMA, seen in Fig. \ref{fig5}.
\par
\par
Since DOS $\rho(\omega)$ is given by the momentum-summation of the spectral weight $A({\bm p},\omega)$ (see Eq. (\ref{eq.8b})), the spectral intensity around $\omega=0$ seems also to be suppressed in the pseudogap regime. However, Figs. \ref{fig7}(a1) and (b1) actually show that, although one can slightly see the coupling of particle ($\omega=\varepsilon_{\bm p}-{\tilde \mu}$) and hole ($\omega=-[\varepsilon_{\bm p}-{\tilde \mu}]$) branches (which is also characteristic of the pseudogap phenomenon associated with pairing fluctuations\cite{Tsuchiya}), such suppression around $\omega=0$ is not clearly seen, because of broad particle and hole dispersions. While this is consistent with the relatively large value of $\rho(\omega=0)$ even in the pseudogap regime, it might become an obstacle in observing the pseudogap by using the photoemission-type experiment\cite{Jin1,Jin2,Jin3,Jin4}. We briefly note that the suppression of the spectral weight around $\omega=0$ is clearly seen in the strong-coupling regime when ${\tilde \mu}<0$, as shown in Fig. \ref{fig7}(c1). 
\par
On the other hand, Figs. \ref{fig7}(a2)-(c2) shows that TMA gives the BCS-state-like spectral weight, where the two sharp peak lines are similar to the Bogoliubov single-particle dispersions in the mean-field BCS state,
\begin{equation}
E_\pm({\bm p})=\pm \sqrt{(\varepsilon_{\bm p}-\mu)^2+|\Delta|^2},
\label{eq12d}
\end{equation}
although the system is in the normal state. 
\par
\begin{figure}[t!]
\begin{center}
\includegraphics[width=7.5cm]{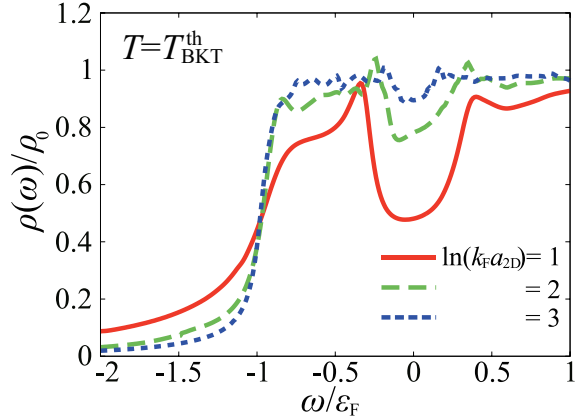}
\caption{(Color online) Calculated DOS $\rho(\omega)$ in SCTMA, in the weak-coupling regime of a two-dimensional Fermi gas at $T_{\rm BKT}^{\rm th}$\cite{Botelho,Tempere}.
}
\label{fig8}       
\end{center}
\end{figure}

From the viewpoint of the recent BKT theory for a two-dimensional Fermi gas which assumes a fixed $|\Delta|$\cite{Botelho,Tempere}, Figs. \ref{fig6} and \ref{fig7}(a2)-(c2) shows that this assumption is justified in TMA, because the (almost) {\it fully-gapped} DOS is indeed realized near $T_{\rm BKT}^{\rm th}$. However, as discussed previously, this gapped structure in TMA is actually filled up to some extent by the quasi-particle lifetime effect in SCTMA (except in the strong-coupling regime where ${\tilde \mu}<0$). This SCTMA result physically means that, not only phase fluctuations, but also {\it amplitude fluctuations} of the superfluid order parameter $\Delta$, are important near $T_{\rm BKT}^{\rm th}$. We also see in Fig. \ref{fig8} that the pseudogap at $T_{\rm BKT}^{\rm th}$ becomes less remarkable, as one approaches the weak-coupling regime.
\par
At present, we have no idea about how to extend the recent BKT theory\cite{Botelho,Tempere} to include amplitude fluctuations. However, since the difference between $T_{\rm BKT}^{\exp}$\cite{Ries} and $T_{\rm BKT}^{\rm th}$\cite{Botelho,Tempere} becomes remarkable in the weak-coupling regime (see Fig. \ref{fig4}), this improvement might be a key to resolve this discrepancy, which remains as our future challenge.  
\par
\par
\section{Assessment of calculated pseudogap from the viewpoint of observed Tan's contact}
\par
In the current stage of cold Fermi gas physics, DOS has not been measured, so that we cannot assess the SCTMA scheme by directly comparing the calculated DOS with experimental data. In this section, however, we demonstrate that we can still do this assessment to some extent, by using the Tan's contact\cite{Tan}, which has recently been measured in a two-dimensional $^{40}$K Fermi gas\cite{2DTan40K}, 
\par
To simply grasp our idea, we first deal with the TMA case, to clarify that it really overestimates the pseudogap size $E_{\rm PG}$. When low-energy pairing fluctuations are strong, the TMA self-energy $\Sigma_{\rm TMA}({\bm p},i\omega_n)$ may be approximated to\cite{Tsuchiya,Chen,note20}
\begin{eqnarray}
\Sigma_{\rm TMA}({\bm p},i\omega_n)
&=&
T\sum_{{\bm q}\nu_n}\Gamma({\bm q},i\nu_n)
G_0({\bm q}-{\bm p},i\nu_n-i\omega_n)
\nonumber
\\
&\simeq&
-\Delta_{\rm PG}^2
G_0(-{\bm p},-i\omega_n).
\label{eq20}
\end{eqnarray}
Here, the particle-particle scattering matrix $\Gamma({\bm q},i\nu_n)$ in TMA is given in Eq. (\ref{eq5}) where the dressed Green's function $G$ in the pair correlation function $\Pi({\bm q},i\nu_n)$ is replaced by the bare one $G_0$. In Eq. (\ref{eq20}),
\begin{equation}
\Delta_{\rm PG}=\sqrt{-T\sum_{{\bm q},i\nu_n}\Gamma({\bm q},i\nu_n)}
\label{eq21}
\end{equation}
is the pseudogap parameter\cite{Chen}. In this so-called static approximation for pairing fluctuations, the TMA single-particle Green's function $G_{\rm TMA}^{\rm static}({\bm p},i\omega_n)$ formally has the same form as the diagonal component of the mean-field BCS Green's function as\cite{Tsuchiya,Schrieffer},
\begin{eqnarray}
G_{\rm TMA}^{\rm static}({\bm p},i\omega_n)=-
{i\omega_n+\xi_{\bm p}
\over
\omega_n^2+\xi_{\bm p}^2+\Delta_{\rm PG}^2}.
\label{eq23G}
\end{eqnarray}
Thus, the resulting density of state has the same form as that in the ordinary BCS state with the gap size $E_{\rm PG}=\Delta_{\rm PG}$. 
\par
\begin{figure}[t!]
\begin{center}
\includegraphics[width=7.5cm]{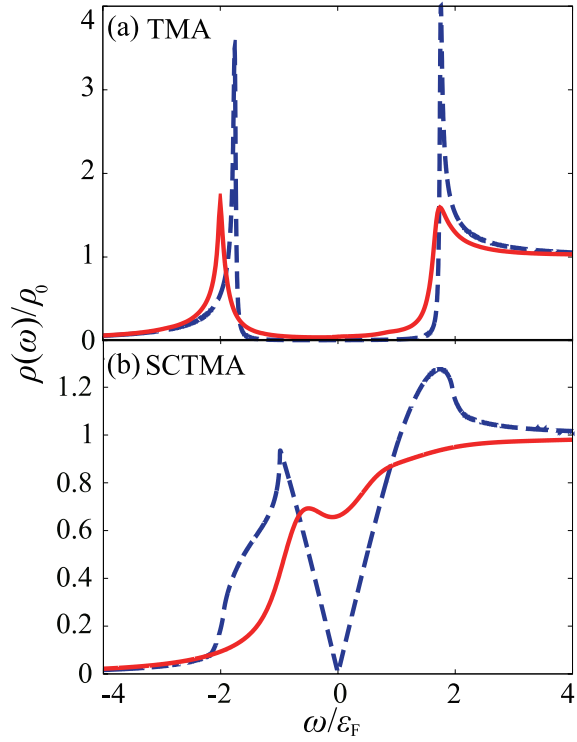}
\caption{(Color online) Calculated DOS in TMA (a), and SCTMA (b). In each figure, the solid line shows the full TMA and SCTMA result, and the dashed line shows the result in the static approximation. We take $\ln(k_{\rm F}a_{\rm 2D})=0.853$ and $T/T_{\rm F}=0.27$ in order to compare our results with the recent experiment on a two-dimensional $^{40}$K Fermi gas\cite{2DTan40K}. In this experiment, the Tan's contact is measured as $C/k^4_{\rm F}=0.223$. TMA and SCTMA give $C/k^4_{\rm F}=0.766$ ($\Delta_{\rm PG}/\varepsilon_{\rm F}=1.75$) and $C/k^4_{\rm F}=0.238$ ($\Delta_{\rm PG}/\varepsilon_{\rm F}=0.975$), respectively.}
\label{fig9}       
\end{center}
\end{figure}
\par
Figure \ref{fig9}(a) compares the ``BCS density of states" obtained from Eq. (\ref{eq23G}) with the full TMA result, where the static approximation is found to well describe the TMA result. Then, noting that the pseudogap parameter $\Delta_{\rm PG}$ in Eq. (\ref{eq21}) is related to the Tan's contact as\cite{Pieri,Houcke}
\begin{equation}
C=m^2\Delta_{\rm PG}^2,
\label{eq22}
\end{equation}
we find that the observable Tan's contact can be used to assess the TMA pseudogap size $E_{\rm PG}$ in Fig. \ref{fig9}(a). Evaluating Eq. (\ref{eq21}) in TMA, we have $\Delta_{\rm PG}/\varepsilon_{\rm F}=1.75$. Although this is consistent with $E_{\rm PG}/\varepsilon_{\rm F}=1.686$ evaluated from the TMA result in Fig. \ref{fig9}(a), it gives $C_{\rm TMA}/k^4_{\rm F}=0.766$, which is much larger than the recent experimental result $C/k^4_{\rm F}=0.223$\cite{2DTan40K} in the same situation as Fig. \ref{fig9}. This disagreement clearly means the overestimation of the pseudogap size in TMA, at least in the case of Fig. \ref{fig9}.
\par
We now proceed to the SCTMA case. In the case of Fig. \ref{fig9}(b), SCTMA gives $\Delta_{\rm PG}/\varepsilon_{\rm F}=0.975$. Substituting this into Eq. (\ref{eq22}), we obtain
\begin{equation}
C_{\rm SCTMA}/k^4_{\rm F}=0.238.
\label{eq23}
\end{equation}
While these values ($\Delta_{\rm PG}$ and $C_{\rm SCTMA}$) are smaller than the TMA results, Eq. (\ref{eq23}) agrees well with the observed value $C/k^4_{\rm F}=0.223$\cite{2DTan40K} in a two-dimensional $^{40}$K Fermi gas. Although one cannot simply relate $\Delta_{\rm PG}$ to the pseudogap size $E_{\rm PG}$ in the case of SCTMA, the smaller pseudogap gap size ($E_{\rm PG}/\varepsilon_{\rm F}=0.46$) in SCTMA (Fig. \ref{fig9}(b)) than that in TMA (Fig. \ref{fig9}(a)) is considered to come from the smaller value $\Delta_{\rm PG}/\varepsilon_{\rm F}=0.975 $ in the former approximation than in the latter ($\Delta_{\rm PG}/\varepsilon_{\rm F}=1.75$). Thus, although we need further analyses to check the validity of SCTMA for the study of the pseudogap phenomenon in a two-dimensional Fermi gas, at least, the calculated pseudogap size in SCTMA does not contradict with the observed value of the Tan's contact $C$. We briefly note that it has recently been shown\cite{Bauer} that SCTMA well explains the observed Tan's contact $C$ in a two-dimensional Fermi gas\cite{2DTan40K} at various interaction strengths.
\par
Before ending this section, we briefly note that the SCTMA single-particle Green's function $G_{\rm SCTMA}^{\rm static}({\bm p},i\omega_n)$ in the static approximation has the form (For the derivation, see the Appendix.),
\begin{equation}
G^{\rm static}_{\rm SCTMA}({\bm p},i\omega_n)=-
{i\omega_n+{\tilde \xi}_{\bm p} \over
\omega_n^2+{\tilde \xi}_{\bm p}^2
+{\tilde \Delta}^2_{\rm PG}({\bm p},i\omega_n)}.
\label{eq24}
\end{equation}
Here, ${\tilde \xi}_{\bm p}=\varepsilon_{\bm p}-{\tilde \mu}$, and
\begin{equation}
{\tilde \Delta}_{\rm PG}({\bm p},i\omega_n)=\Delta_{\rm PG}
\sqrt{
2
\over 
1+\sqrt{1+{4\Delta^2_{\rm PG} \over \omega_n^2+{\tilde \xi}_{\bm p}^2}}
},
\label{eq25}
\end{equation}
where the pseudogap parameter $\Delta_{\rm PG}$ is also given in Eq. (\ref{eq21}), but the SCTMA particle-particle scattering matrix $\Gamma({\bm q},i\nu_n)$ in Eq. (\ref{eq5}) is now used. 
\par
Comparing Eq. (\ref{eq24}) with Eq. (\ref{eq23G}), one finds that the pseudogap parameter $\Delta_{\rm PG}$ in TMA is replaced by the momentum- and energy-dependent parameter ${\tilde \Delta}_{\rm PG}({\bm p},i\omega_n)$ in Eq. (\ref{eq25}). In this regard, we note that this (${\bm p},\omega_n$)-dependence comes from the self-energy corrections in the dressed Green's function $G$ in the outer loop of the SCTMA self-energy in Fig. \ref{fig1}. In this sense, effects of quasi-particle lifetime are still partially taken into account in this static approximation for SCTMA. Indeed, this $({\bm p},\omega_n)$-dependent pseudogap parameter ${\tilde \Delta}_{\rm PG}({\bm p},i\omega_n)$ gives the partially-filled pseudogap structure (except at $\omega=0$), as shown in Fig. \ref{fig9}(b). However, Fig. \ref{fig9}(b) also shows that this static approximation overestimates the pseudogap size, indicating that it underestimates effects of quasi-particle lifetime on pairing fluctuations described by the particle-particle scattering matrix $\Gamma({\bm q},i\nu_n)$. 
\par
\par
\section{Summary}
\par
To summarize, we have discussed pseudogap phenomena in a two-dimensional Fermi gas. Including pairing fluctuations within the framework of the self-consistent $T$-matrix approximation (SCTMA), we have calculated single-particle density of states $\rho(\omega)$ (DOS), as well as the single-particle spectral weight $A({\bm p},\omega)$, in the normal state. Determining the pseudogap temperature below which a dip appears in $\rho(\omega\sim 0)$, we have identify the region where the pseudogap appears (pseudogap regime) in the phase diagram with respect to the interaction strength and the temperature. We have also examined the detailed pseudogap structure near the BKT transition temperature $T_{\rm BKT}^{\rm th}$ which has been obtained by the recent BKT theory\cite{Botelho,Tempere} assuming a fixed magnitude of the superfluid order parameter $|\Delta|$, to assess the validity of this assumption.
\par
We showed that the pseudogap regime widely exists in the weak-coupling regime of a two-dimensional Fermi gas, compared to the three-dimensional case. This is an expected result, because the two-dimensionality of the system should enhance pairing fluctuations. However, we also found that the pseudogap temperature $T^*$, as well as the pseudogap size $E_{\rm PG}$, are not so sensitive to the interaction strength in the weak and intermediate coupling regime. This is because the increase of the pairing interaction shortens the lifetime of Fermi quasi-particles, which tends to suppress the enhancement of pairing fluctuations by the same pairing interaction. Indeed, when the quasi-particle lifetime effects on pairing fluctuations are completely ignored in the non-selfconsistent $T$-matrix approximation (TMA), the pseudogap size $E_{\rm PG}$ grows more remarkably with increasing the interaction strength than the SCTMA result.
\par
For the assessment of the recent BKT theory assuming a fixed amplitude $|\Delta|$ of the superfluid order parameter\cite{Botelho,Tempere}, our results show the importance of amplitude fluctuations near the BKT transition temperature $T_{\rm BKT}^{\rm th}$ which is obtained by this BKT theory. That is, the calculated pseudogapped DOS at $T_{\rm BKT}^{\rm th}$ in SCTMA still has sizable intensity around $\omega=0$, especially in the weak-coupling regime. Since TMA gives the BCS-state-like almost fully gapped DOS at $T_{\rm BKT}^{\rm th}$ (which justifies the assumption of fixed $|\Delta|$), the above-mentioned quasi-particle lifetime is found to also play an important role in this problem. Since $T_{\rm BKT}^{\rm th}$ is known to deviate from the observed BKT transition temperature $T_{\rm BKT}^{\rm exp}$\cite{Ries} in the weak-coupling regime, it is an interesting future challenge to improve this BKT theory to include amplitude fluctuations, to see to what extent this improvement can resolve this discrepancy.
\par
At present, DOS has not been measured in an ultracold Fermi gas, so that we cannot directly compare our results with experimental data. However, we pointed out that the Tan's contact (which has recently been observed in a two-dimensional Fermi gas\cite{2DTan40K}) can be used to assess our theoretical results to some extent. Using this approach, we showed that, while the TMA overestimates the pseudogap size, the pseudogap size in SCTMA does not contradict with the observed value of the Tan's contact $C$. Of course, we need further analyses for the validity of SCTMA for the study of pseudogap phenomena in a two-dimensional Fermi gas.
\par
While the wide pseudogap regime is an advantage of a two-dimensional Fermi gas in studying the pseudogap phenomenon, we still need to overcome the smearing of pseudogap structure in $\rho(\omega)$ and $A({\bm p},\omega)$ by quasi-particle lifetime, which also comes from the pairing interaction. In the current stage of cold Fermi gas physics, the photoemission-type experiment would be the most effective to directly measure single-particle properties of an ultracold Fermi gas. Thus, it is also an interesting future problem to theoretically evaluate the photoemission spectrum in the pseudogap regime of a two-dimensional Fermi gas, by using the present SCTMA formalism. Since the pseudogap is a key phenomenon in understanding strong-coupling properties of an ultracold Fermi gas, as well as in constructing BKT theory for a two-dimensional Fermi gas, our results would be useful for the studies of these crucial current topics in cold Fermi gas physics.
\par

\begin{acknowledgment}

\par
 We thank H. Tajima, P. van Wyk, D. Kharga, and D. Kagamihara for discussions. This work was supported by the KiPAS project in Keio university. M.M. was supported by a KLL Ph.D. Program Research Grant. D.I. was supported by Grant-in-Aid for Young Scientists (B) (No. JP16K17773) from JSPS. R.H. was supported by Grant-in-Aid for JSPS fellows. Y.O. was supported by Grand-in-Aid for Scientific Research from MEXT and JSPS in Japan (No. JP15K00178, No. JP15H00840, No. JP16K05503).
\par
\end{acknowledgment}
\par
\appendix
\section{Static approximation for SCTMA}
\par
Assuming that low-energy pairing fluctuations are strong, we approximate the SCTMA self-energy in Eq. (\ref{eq4}) to
\begin{equation}
\Sigma({\bm p},i\omega_n)\simeq -\Delta_{\rm PG}^2G(-{\bm p},-i\omega_n)+\Sigma_{\rm HF},
\label{eq.app1}
\end{equation}
where the pseudogap parameter $\Delta_{\rm PG}$ is given in Eq. (\ref{eq21}). In Eq. (\ref{eq.app1}), we also retain a constant Hartree term $\Sigma_{\rm HF}$, in order to effectively include the difference between $\mu$ and ${\tilde \mu}=\mu+\Sigma_{\rm HF}$ shown in Fig. \ref{fig2}. Substituting Eq. (\ref{eq.app1}) into Eq. (\ref{eq3}), we have
\begin{eqnarray}
G({\bm p},i\omega_n)
&=&
{1 \over i\omega_n-{\tilde \xi}_{\bm p}+
\Delta_{\rm PG}^2G(-{\bm p},-i\omega_n)}
\nonumber
\\
&=&
{1 
\over 
i\omega_n-{\tilde \xi}_{\bm p}
+
{\displaystyle \Delta_{\rm PG}^2
\over
\displaystyle
-i\omega_n-{\tilde \xi}_{-{\bm p}}+\Delta_{\rm PG}^2G({\bm p},i\omega_n)}
}.
\nonumber
\\
\label{eq.app2}
\end{eqnarray}
Here, the Hartree energy $\Sigma_{\rm HF}$ is absorbed into ${\tilde \xi}_{\bm p}=\varepsilon_{\bm p}-{\tilde \mu}$. In obtaining the last expression, we have substituted the first expression into $G(-{\bm p},-i\omega_n)$ in the denominator of this expression. Solving Eq. (\ref{eq.app2}) with respect to $G({\bm p},i\omega_n)$ we obtain Eq. (\ref{eq24}).
\par


\begin{thebibliography}{99}
\bibitem{Timmermans} E. Timmermans, K. Furuya, P. W. Milonni, and A. K. Kerman, Phys. Lett. A {\bf 285}, 228 (2001). 
\bibitem{Chin} C. Chin, R. Grimm, P. Julienne, and E. Tiesinga, Rev. Mod. Phys. {\bf 82}, 1225 (2010).
\bibitem{Morsch} O. Morsch, and M. Oberthaler, Rev. Mod. Phys. {\bf 78}, 179 (2006). 
\bibitem{Bloch} I. Bloch, J. Dalibard, and W. Zwerger, Rev. Mod. Phys. {\bf 80}, 885 (2008).
\bibitem{Perali} A. Perali, P. Pieri, G. C. Strinati, and C. Castellani, Phys. Rev. B {\bf 66}, 024510 (2002).
\bibitem{Levin} Y. He, Q. Chen, and K. Levin, Phys. Rev. A {\bf 72}, 011602(R) (2005).
\bibitem{Tsuchiya} S. Tsuchiya, R. Watanabe, and Y. Ohashi, Phys. Rev. A {\bf 80}, 033613 (2009).
\bibitem{Watanabe1}R. Watanabe, S. Tsuchiya, and Y. Ohashi, Phys. Rev. A {\bf 82}, 043630 (2010).
\bibitem{Botelho} S. S. Botelho and C. A. R. S$\acute{{\rm a}}$ de Melo, Phys. Rev. Lett. {\bf 96}, 040404 (2006).
\bibitem{Tempere} J. Tempere, S. N. Klimin, and J. T. Devreese,	{Phys. Rev.} A {\bf 79}, 053637 (2009). 
\bibitem{Pietilla} V. Pietil$\ddot{\rm a}$, Phys. Rev. A {\bf 86}, 023608 (2012).
\bibitem{Watanabe} R. Watanabe, S. Tsuchiya, and Y. Ohashi, Phys. Rev. A {\bf 88}, 013637 (2013).
\bibitem{Bauer} M. Bauer, M. M. Parish, T. Enss, Phys. Rev. Lett. {\bf 112}, 135302 (2014).
\bibitem{Mulkerin} B. C. Mulkerin, K. Fenech, P. Dyke, C. J. Vale, X. Liu, and H. Hu, Phys. Rev. A {\bf 92}, 063636 (2015).
\bibitem{Marsiglio}  F. Marsiglio, P. Pieri, A. Perali, F. Palestini, and G. C. Strinati, Phys. Rev. B {\bf 91}, 054509 (2015).
\bibitem{Matsumoto1}  M. Matsumoto, D. Inotani, and Y. Ohashi, Phys. Rev. A {\bf 93}, 013619 (2016).
\bibitem{Salasnich} L. Salasnich, and G. Bighin, J. Supercond. Nov. Magn., {\bf 29}, 3103 (2016).
\bibitem{Mulkerin1} B. C. Mulkerin, L. He, P. Dyke, C. J. Vale, X. Liu, H. Hu , arXiv e-prints, 1702.07091 (2017).		
\bibitem{Martiyanov} K. Martiyanov, V. Makhalov, and A. Turlapov, Phys. Rev. Lett. {\bf 105}, 030404 (2010).
\bibitem{Feld} M. Feld, B. Fr\"ohlich, E. Vogt, M. Koschorreck, and M. K\"ohl, Nature {\bf 480}, 75 (2011).
\bibitem{Frohlich} B. Fr\"ohlich, M. Feld, E. Vogt, M. Koschorreck, W. Zwerger, and M. K\"ohl, Phys. Rev. Lett. {\bf 106}, 105301 (2011).
\bibitem{Sommer} A. T. Sommer, L. W. Cheuk, M. J. H. Ku, W. S. Bakr, and M. W. Zwierlein, Phys. Rev. Lett. {\bf 108}, 045302 (2012).
\bibitem{Makhalov} V. Makhalov, K. Martiyanov, and A. Turlapov,	Phys. Rev. Lett. {\bf 112}, 045301 (2014).
\bibitem{Fenech} K. Fenech, P. Dyke, T. Peppler, M. G. Lingham, S. Hoinka, H. Hu, and C. J. Vale, Phys. Rev. Lett. {\bf 116}, 045302 (2016).
\bibitem{Ries} M. G. Ries, A. N. Wenz, G. Z\"urn, L. Bayha, I. Boettcher, D. Kedar, P. A. Murthy, M. Neidig, T. Lompe, and S. Jochim, Phys. Rev. Lett. {\bf 114}, 230401 (2015).
\bibitem{Murthy2} P. A. Murthy, I. Boettcher, L. Bayha, M. Holzmann, D. Kedar, M. Neidig, M. G. Ries, A. N. Wenz, G. Z\"urn, and S. Jochim, Phys. Rev. Lett. {\bf 115}, 010401 (2015).
\bibitem{Berezinskii} V. L. Berezinskii, Sov. Phys. JETP {\bf 32}, 493 (1971).
\bibitem{Berezinskii1} V. L. Berezinskii, Sov. Phys. JETP {\bf 34}, 610 (1972).
\bibitem{Kosterlitz} J. M. Kosterlitz and D. J. Thouless, J. Phys. C {\bf 6}, 1181 (1973). 
\bibitem{Kosterlitz1} J. M. Kosterlitz, J. Phys. C {\bf 7}, 1046 (1974).

\bibitem{Regal} C. A. Regal, M. Greiner, and D. S. Jin, Phys. Rev. Lett. \textbf{92}, 040403 (2004).
\bibitem{Zwierlein} M. W. Zwierlein, C. A. Stan, C. H. Schunck, S. M. F. Raupach, A. J. Kerman, and W. Ketterle, Phys. Rev. Lett. \textbf{92}, 120403 (2004).
\bibitem{Kinast}J. Kinast, S. L. Hemmer, M. E. Gehm, A. Turlapov, and J. E. Thomas, Phys. Rev. Lett. \textbf{92}, 150402 (2004).
\bibitem{Bartenstein} M. Bartenstein, A. Altmeyer, S. Riedl, S. Jochim, C. Chin, J. Hecker Denschlag, and R. Grimm, Phys. Rev. Lett. \textbf{92}, 203201 (2004).
\bibitem{Ketterle} W. Ketterle and M. W. Zwierlein, in {\it Proceedings of the International School of Physics Enrico Fermi, Course CLXIV}, edited by M. Inguscio, W. Ketterle, and C. Salomon (IOS, Amsterdam, 2008).
\bibitem{Zwerger} W. Zwerger (Ed), {\it The BCS-BEC Crossover and the Unitary Fermi Gas} (Springer, Heidelberg, 2012).
\bibitem{Eagles} D. M. Eagles, Phys. Rev. {\bf 186}, 456 (1969).
\bibitem{Leggett} A. J. Leggett, in {\it Modern Trends in the Theory of Condensed Matter}, edited by A. Pekalski and J. Przystawa (Springer, Berlin, 1980), p.14.
\bibitem{NSR} P. Nozi$\grave{{\rm e}}$res and S. Schmitt-Rink, J. Low Temp. Phys. {\bf 59}, 195 (1985).
\bibitem{Melo} C. A. R. S$\acute{{\rm a}}$ de Melo, M. Randeria, and J. R. Engelbrecht, Phys. Rev. Lett. {\bf 71}, 3202 (1993).
\bibitem{Randeria} M. Randeria, in {\it Bose-Einstein Condensation}, edited by A. Griffin, D. W. Snoke, and S. Stringari, (Cambridge University Press, New York, 1995), p. 355.
\bibitem{Ohashi2002} Y. Ohashi and A. Griffin, Phys. Rev. Lett. {\bf 89}, 130402 (2002).
\bibitem{Levin2005} Q. Chen, J. Stajic, S. Tan, and K. Levin, Phys. Rep. {\bf 412}, 1 (2005).
\bibitem{Gurarie} V.  Gurarie, and L. Radzihovsky, Ann. Phys. {\bf 332}, 2 (2007).
\bibitem{Giorgini} S. Giorgini, S. Pitaevskii, and S. Stringari, Rev. Mod. Phys. \textbf{80}, 1215 (2008).
\bibitem{Chen} Q. J. Chen and K. Levin, Phys. Rev. Lett. {\bf 102}, 190402 (2009). 
\bibitem{Hu} H. Hu, X.-J. Liu, P. D. Drummond, and H. Dong, Phys. Rev. Lett. {\bf 104}, 240407 (2010). 
\bibitem{Mueller} E. J. Mueller, Phys. Rev. A {\bf 83}, 053623 (2011). 
\bibitem{Bulgac} P. Magierski, G. Wlazlowski, and A. Bulgac, Phys. Rev. Lett. {\bf 107}, 145304 (2011). 
\bibitem{Jarrell} S-Q. Su, D. E. Sheehy, J. Moreno, and M. Jarrell, Phys. Rev. A {\bf 81}, 051604(R) (2010). 
\bibitem{Bulgac2} G. Wlaz\l owski P. Magierski, J. E. Drut, A. Bulgac, and K. J. Roche, Phys. Rev. Lett. {\bf 110}, 090401 (2013).
\bibitem{Ota} M. Ota, H. Tajima, R. Hanai, D. Inotani, and Y. Ohashi, Phys. Rev. A {\bf 95}, 053623 (2017).
\bibitem{Jin1} J. T. Stewart, J. P. Gaebler, and D. S. Jin, Nature (London) {\bf 454}, 744 (2008). 
\bibitem{Jin2} J. P. Gaebler, J. T. Stewart, T. E. Drake, D. S. Jin, A. Perali, P. Pieri, and G. C. Strinati, Nat. Phys. {\bf 6}, 569 (2010). 
\bibitem{Jin3} A. Perali, F. Palestini, P. Pieri, G. C. Strinati, J. T. Stewart, J. P. Gaebler, T. E. Drake, and D.S.Jin, Phys. Rev. Lett. {\bf 106}, 060402 (2011). 
\bibitem{Jin4} Y. Sagi, T. E. Drake, R. Paudel, R. Chapurin, and D. S. Jin, Phys. Rev. Lett. {\bf 114}, 075301 (2015).
\bibitem{Salomon1} S. Nascimb\`ene, N. Navon, K. J. Jiang, F. Chevy, and C. Salomon, Nature (London) {\bf 463}, 1057 (2010). 
\bibitem{Salomon2} S. Nascimb\`ene, N. Navon, S. Pilati, F. Chevy, S. Giorgini, A. Georges, and C. Salomon, Phys. Rev. Lett. {\bf 106}, 215303 (2011).
\bibitem{Randeria1992} M. Randeria, N. Trivedi, A. Moreo, and R. T. Scalettar, Phys. Rev. Lett. {\bf 69}, 2001 (1992).
\bibitem{Singer} J. M. Singer, M. H. Pedersen, T. Schneider, H. Beck, and H.-G. Matuttis, Phys. Rev. B {\bf 54}, 1286 (1996).
\bibitem{Rohe} D. Rohe and W. Metzner, Phys. Rev. B {\bf 63}, 224509 (2001).
\bibitem{Yanase}  Y. Yanase and K. Yamada, J. Phys. Soc. Jpn. {\bf 70}, 1659 (2001).
\bibitem{Kadowaki} Ch. Renner, B. Revaz, J.-Y. Genoud, K. Kadowaki, and \O. Fischer, Phys. Rev. Lett. {\bf 80}, 149 (1998).
\bibitem{Shen} A. Damascelli, Z. Hussain, and Z.-X. Shen, Rev. Mod. Phys. {\bf 75}, 473 (2003).
\bibitem{Fischer} \O. Fischer, M. Kugler, I. Maggio-Aprile, and C. Berthod, Rev. Mod. Phys. {\bf 79}, 353 (2007).
\bibitem{Hohenberg} P. Hohenberg, Phy. Rev. {\bf 158}, 383 (1967).
\bibitem{Mermin} N. D. Mermin, and H. Wagner, Phys. Rev. Lett. {\bf 17}, 1133 (1966). 
\bibitem{Haussmann} R. Haussmann, Z. Phys. B: Condens. Matter {\bf 91}, 291(1993).
\bibitem{Haussmann2} R. Haussmann, W. Rantner, S. Cerrito, and W. Zwerger, Phys. Rev. A {\bf 75}, 023610 (2007).
\bibitem{Nelson} D. R. Nelson, J. M. Kosterlitz, Phys. Rev. Lett. {\bf 39} 1201 (1977).  
\bibitem{MatsumotoJLTP} M. Matsumoto, R. Hanai, D. Inotani, Y. Ohashi, J. Low Temp. Phys. {\bf 187}, 668 (2017).

\bibitem{Tan} S. Tan, Ann. Phys. {\bf 323}, 2971 (2008).
\bibitem{2DTan40K} B. Fr\"ohlich, M. Feld, E. Vogt, M. Koschorreck, M. K\"ohl, C. Berthod, and T. Giamarchi, Phys. Rev. Lett. {\bf 109}, 130403 (2012).
\bibitem{Morgan} S. A. Morgan, M. D. Lee, and K. Burnett, Phys. Rev. A {\bf 65}, 022706 (2002).
\bibitem{Petrov}D. S. Petrov and G. V. Shlyapnikov, Phys. Rev. A {\bf 64}, 012706 (2001). 
\bibitem{Levinsen}J. Levinsen and M. M. Parish, Annu. Rev. Cold Atoms and Molecules {\bf 3}, 1. 
\bibitem{noteEb} We note that another expression for the binding energy, $E_{\rm bind}=2/(C^2m^2\widetilde{a}_{\rm 2D}^{2})$, is also used\cite{Levinsen}, where $C=e^{\gamma}$ with $\gamma=0.577$ being the Euler constant. This difference, however, simply comes from the definition for the two-dimensional s-wave scattering length $\widetilde{a}_{\rm 2D}$, which is related to our $a_{\rm 2D}$ as $a_{\rm 2D}=C\widetilde{a}_{\rm 2D}/\sqrt{2}$. 
\bibitem{Vidberg} H. J. Vidberg, and J. W. Serene, J. Low Temp. Phys. {\bf 29}, 179 (1977).
\bibitem{Thouless} D. J. Thouless, Ann. Phys. (N. Y.) {\bf 10}, 553 (1960).
\bibitem{Varma} S. Schmitt-Rink, C. M. Varma, and A. E. Ruckenstein, Phys. Rev. Lett. {\bf 63}, 445 (1989).
\bibitem{Tokumitu} A. Tokumitu, K. Miyake, and K. Yamada Phys. Rev. B. {\bf 47}, 11988 (1993).

\bibitem{note} When single-particle excitations are completely dominated by molecular dissociation, the width of the energy gap in the density of states must equal $E_{\rm bind}$ in Eq. (\ref{eq2b}). Regarding this, one has $E_{\rm bind}=14.8\varepsilon_{\rm F}$ when $\ln(k_{\rm F}a_{\rm 2D})=-1$. This value is still different from the gap width ($\sim 4\varepsilon_{\rm F}$) in Fig. \ref{fig3}(c) at $T/T_{\rm F}=0.3$. This means that the system at this interaction strength is still not deep inside the strong-coupling regime.

\bibitem{noteZ} In our numerical results, while the pseudogap is always accompanied by a peak structure at the lower pseudogap edge, it sometimes does not exhibit a BCS-state-like double-peak structure at the upper and lower pseudogap edges. Because of this, we take this definition for the pseudogap size in this paper.
\bibitem{noteW} Of course, effects of quasi-particle lifetime also comes from the dressed Green's function $G$ in the outer loop of the self-energy in Fig. \ref{fig1}. As discussed in Sec. IV, this contributes to filling up the pseudogap (see also Fig. \ref{fig9}(b)).

\bibitem{note20} We briefly note that a better approximation is to include a constant Hartree shift $\Sigma_{\rm HF}$, as $\Sigma_{\rm TMA}({\bm p},i\omega_n)\simeq \Sigma_{\rm HF}-\Delta_{\rm PG}^2G_0(-{\bm p},-i\omega_n)$\cite{Tsuchiya}. However, we do not include $\Sigma_{\rm HF}$ to simplify our discussion in this paper. We take into account this term in considering SCTMA (see the Appendix).
\bibitem{Schrieffer} J. R. Schrieffer, {\it Theory of Superconductivity} (Addison-Wesley, New York, 1964).
\bibitem{Pieri}P. Pieri, A. Perali, and G. C. Strinati, Nat. Phys. {\bf 5}, 736 (2009).
\bibitem{Houcke} K. Van Houcke, F. Werner, E. V. Kozik, N. V. Prokof'ev, and B. V. Svistunov, arXiv:1303.6245.
\end{thebibliography}
\end{document}